\documentstyle[aps,twocolumn]{revtex}

\begin{document}
\draft
\def\be{\begin{equation}}
\def\ee{\end{equation}}
\def\beq{\begin{eqnarray}}
\def\eeq{\end{eqnarray}}
\input{psfig}
\wideabs{
\title
{
Orientation of the Stripe Formed by the Two-Dimensional Electrons\\
 in Higher Landau Levels
}
\author
{
Daijiro Yoshioka
}
\address
{
Department of Basic Science, The University of Tokyo\\
Komaba, Meguro-ku, Tokyo 153-8902\\
}
\date{June 27,2001}
\maketitle
\begin{abstract}
Effect of periodic potential on the stripe phase realized in the higher Landau levels is investigated by the Hartree-Fock approximation.
The period of the potential is chosen to be two to six times of the fundamental period of the stripe phase.
It is found that the stripe aligns perpendicularly to the external potential in contrast to a naive expectation and hydrodynamic theory.
Charge modulation towards the Wigner crystallization along the stripe is essential for the present unexpected new result.
\end{abstract}

\pacs
}

When the two-dimensional electrons are in a strong magnetic field such that the lowest Landau level is partially filled, the fractional quantum Hall effect or the composite fermion Fermi liquid state is realized at low temperatures depending on the filling factor.
On the other hand, it is almost established experimentally~\cite{Lill,Du,Coop} and theoretically~\cite{Kou1,Kou2,Rez1,shibata} that the stripe phase or the unidirectional CDW state is formed in ultra-high mobility samples at low temperature when the third lowest or higher Landau levels is nearly half filled.
The orientation of the stripe is fixed relative to the crystal axis in the perpendicular field, although in-plane magnetic field can affect it.~\cite{Pan,Lill1,Jung}
Hence, the longitudinal resistivity is much larger in the $[1,\bar 1,0]$ direction than in the [1,1,0] direction.
According to a theoretical calculation of the conductivity tensor of the stripe phase,~\cite{Mac} this indicates that the stripe runs along the [1,1,0] direction.

The reason why the stripe has preferred orientation has not been known.
It is hard to believe that the crystal structure affects the orientation of the stripe.
In the experiments the magnetic field is a few teslas, so the wave function of each electron in the third Landau level is spread over more than several hundred angstroms.
So details of the crystal lattice structure will be averaged out.
On the other hand, investigation of the surface morphology of high mobility samples shows structure running along the $[1,\bar 1,0]$ direction.~\cite{Orme,Will}
It is possible that this structure orients the stripe.

To see if this morphology is relevant or not, we consider effect of a periodic potential of wave vector ${\vec Q}_{\rm p}$ on the stripes of fundamental wave vector ${\vec Q}_{\rm s}$ in this letter.
We calculate the ground state energy in two configurations: in one of the configuration ${\vec Q}_{\rm p}$ and ${\vec Q}_{\rm s}$ are parallel to each other and in the other they are perpendicular.
Naively, it seems that the parallel configuration has lower energy.
As a matter of fact, if the sizes of the wave vectors are the same, the lowest energy is achieved when they are parallel to each other.
Moreover, Luttinger liquid theory, or hydrodynamic theory,~\cite{Mac,Fog,Lop} which is considered to be valid in the long wavelength, low energy limit, favors parallel orientation.
This is because in the perpendicular orientation, only one type of deformation of the stripe is caused.
It is modulation of the width of the stripes.
On the other hand, in the parallel orientation, two types of the deformation of the stripe are caused at the same time.
The additional deformation is the displacement of the stripes.
This additional coupling to the displacement brings more energy gain in the parallel orientation.
We will come back to this point later.

Now if ${\vec Q}_{\rm p}$ and ${\vec Q}_{\rm s}$ are parallel, we are in trouble.
In the experiments, the wave vector of potential modulation ${\vec Q}_{\rm p}$ should be in the [1,1,0] direction, and the resistivity is lower in the same direction, which indicates ${\vec Q}_{\rm s}$ is in the $[1,\bar 1,0]$ direction.
In order to resolve this paradox we present in this letter results of microscopic Hartree-Fock calculation of the stripe phase in the external potential modulation.
The calculation is done for the third lowest Landau level at half-filling.
Our results show that contrary to the long wavelength theory the wave vectors of the stripe and the potential are orthogonal to each other at least when the wavelength of the potential modulation is a few times of the period of the stripe.

Our Hamiltonian consists of two parts:
The Coulomb interaction part for the third Landau level is given as follows:
\be
H_0 = \frac{1}{L^2} \sum_{\vec q} v(q) [L_2(\frac{1}{2}q^2l^2)]^2 \rho({\vec q}) \rho(-{\vec q}),
\label{eq:1}
\ee
\be
v(q) = \frac{e^2}{2\epsilon q},
\label{eq:2}
\ee
\be
\rho({\vec q})=\sum_X \exp[-{\rm i}q_xX-\frac{(ql)^2}{4}] a_{X_+}^\dagger a_{X_-},
\ee
\be
X_{\pm}=X\pm \frac{1}{2}l^2q_y,~~0 \le X=\frac{2\pi l^2}{L}j \le L.
\ee
In these equations $\rho({\vec q})$ is the density operator of the guiding center, $L_n(x)$ is the Laguerre polynomial, $l=\sqrt{\hbar/eB}$ is the magnetic length, and $L \to \infty$ is the linear dimension of the system.
The other part is the interaction term with the external potential,
\be
H_1 = \frac{1}{2} V_{\rm ext} \sum_{{\vec q}=\pm {\vec Q}_{\rm p}} \rho({\vec q}) L_2(\frac{1}{2}q^2l^2) \exp({\rm i}{\vec q}{\vec r}_0),
\label{eq:5}
\ee
where $V_{\rm ext}$ is the strength of the potential, and ${\vec r}_0$ is the origin of the potential.

In the Hartree-Fock theory~\cite{YH,YL} we introduce order parameters
\be
\Delta({\vec Q}) = \frac{2\pi l^2}{L^2}\sum_X \langle a_{X_+}^\dagger a_{X_-}\rangle \exp(-{\rm i}Q_xX),
\ee
where $
X_\pm = X \pm ({1}/{2})l^2Q_y$.
The wave vectors of the order parameters are expressed as
\be
{\vec Q}= (jQ_{x}^0, kQ_y^0),
\ee
with $j$ and $k$ being integers.
We recognize a CDW state as the stripe phase when the order parameters with ${\vec Q}=\pm {\vec Q}_s$ are dominant.

In the present calculation we generalized the procedure detailed in ref.\cite{YL} to the case of rectangular symmetry, $Q_x^0 \ne Q_y^0$.
Then when
\be
NQ_x^0Q_y^0 l^2 = 2M\pi,
\ee
with $N$ and $M$ being integers, the Landau level splits into $N$ Hofstadter bands, and we fill half of the bands to get the filling factor $\nu=1/2$.
First we searched the optimum stripe state by changing the integers $N$, $M$ and the ratio $Q_x^0/Q_y^0$.
As a result we found that the choice $Q_x^0=Q_y^0 \equiv Q_{\rm s}=\sqrt{\pi/3}/l$, $N=6$, and $M=1$ gives almost the lowest energy stripe state, the energy being 0.30969$e^2/4\pi\epsilon l$ per electron.
In this calculation we have retained the order parameters with wave vector ${\vec Q}$ such that $|Q|l \le Q_{\rm max}l = \sqrt{200}$.
The Yoshioka-Lee's order parameter sum rule,~\cite{YL}
\be
\nu=\sum_{\vec Q}|\Delta({\vec Q})|^2,
\ee
is satisfied by 99.994\%, which indicates that the contribution from the discarded order parameters is negligible.
The obtained stripe state is degenerated in the direction of the fundamental wave vector as expected.
Namely, we obtained ground state where the fundamental wave vector of the stripe ${\vec Q}_{\rm s}$ is either $(Q_{\rm s},0)$ or $(0,Q_{\rm s})$.
The density along the stripe is modulated slightly with wave number $3Q_{\rm s}$ corresponding to the linear alignment of electrons.
Namely, the unidirectional CDW state has ``$2k_{\rm F}$ instability'' towards the Wigner crystallization, and energy gap is created at the Fermi level as has been first pointed out by Yoshioka and Fukuyama.~\cite{YH}
However, the modulation is not large enough to destroy the topology of the stripe.~\cite{Cote}

Next we take into account the external potential.
We consider external potential with wave vector ${\vec Q}_p=(Q_p,0)$.
We apply this potential to the stripe state that has ${\vec Q}_{\rm s}= (Q_{\rm s},0)$ or $(0,Q_{\rm s})$, respectively.
The wave number of the potential $Q_{\rm p}$ is chosen to be $Q_{\rm s}/m$, where $m$ is an integer.
Then we calculate the order parameters $\Delta(jQ_{\rm p},kQ_{\rm s})$ with $j$ and $k$ integers self-consistently, and obtain the ground state energy in the periodic potential.
Since $Q_x^0$ is decreased to $Q_{\rm p}$, $N$ becomes $6m$.
Namely, the energy spectrum consists of $6m$ bands in this case.
As $m$ becomes larger so does both the size of the matrix we must diagonalize and the number of order parameters;
it takes longer time to obtain the self-consistent solution.
Therefore, we calculated only up to $m=6$.
In the course of the calculation all the order parameters whose wave vector is smaller than $10/l$ is retained except for $m=6$, for which $Q_{\rm max}=\sqrt{40}/l$.
We have checked at smaller $m$ that $Q_{\rm max}=\sqrt{40}/l$ is enough: although the ground state energy changes as we decrease $Q_{\rm max}$, the energy gain in the external potential changes only slightly.
At $m=2$ the difference between the $Q_{\rm max}=\sqrt{200}/l$ and $\sqrt{40}/l$ cases is largest and about 2.5\%, however, at $m=3$ it is less than 1\%.
For $m=4$ and 5 the difference between the $Q_{\rm max}=10/l$ and $\sqrt{40}/l$ cases is much smaller.

The result for the case of $Q_{\rm p}=Q_{\rm s}$ is trivial.
Once the external potential becomes finite, we obtain only a self-consistent solution where the fundamental wave vector of the stripe coincides with that of the external potential.
In this case the ground state energy lowers almost proportional to $V_{\rm ext}$.
We will not discuss this case any more.

For $2 \le m \le 6$ we have obtained two different self-consistent solutions.
Namely, in one solution ${\vec Q}_{\rm s}$ is perpendicular to ${\vec Q}_{\rm p}$, and in the other they are parallel.
The energy gain depends on the origin of the external potential, so optimized position is chosen.
In all the cases the energy gain is almost proportional to $V_{\rm ext}^2$, as long as $V_{\rm ext}$ is smaller than the Hartree-Fock self-consistent potential of the CDW, $\simeq 0.3e^2/4\pi\epsilon l$.
We express the difference as follows:
\be
\delta E \equiv E(V_{\rm ext}) -E(0) = -\frac{1}{2} \varepsilon V_{\rm ext}^2 + O(V_{\rm ext}^3),
\ee
where $E(V_{\rm ext})$ is the ground state energy per electron in $V_{\rm ext}$.
The proportionality constant $\varepsilon$ is given in Table.1.
As shown there the most stable configuration is that where the wave vectors are perpendicular to each other.
This result is opposite to that of the long wavelength theory.

\begin{figure}
\psfig{figure=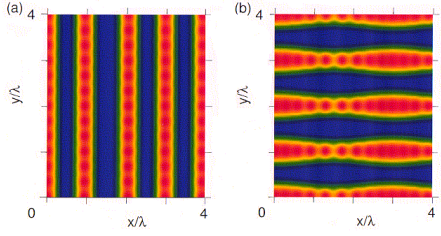,width=8.5cm}
\caption{The real space CDW pattern in the external potential.
In this figure $Q_{\rm p}=Q_{\rm s}/3$ and $V_{\rm ext}=0.2e^2/4\pi\epsilon l$.
The Fourier transform of the $\langle \rho({\vec q}) \rangle$ is shown for an area of linear dimension of four times of the stripe period $\lambda=2\pi/Q_{\rm s}$.  The brighter region shows higher density.  (a) for the case of parallel configuration, and (b) for the perpendicular configuration.}
\label{fig:1}
\end{figure}
The pattern of the stripe in the potential is shown in Fig.1 for the case of $m=3$.
When $Q_{\rm p} \perp Q_{\rm s}$, there is only width modulation.~\cite{note2}
On the other hand, the displacement of the stripe is the main deformation in the parallel case.
For different choice of the origin of the potential, the width of each stripe also changes.
Actually, there is the electron-hole symmetry, so the displacement of the electron stripe means width change of the hole-stripe, and these two patterns give the same energy.

In order to extrapolate our result to longer wavelength, we calculated what we call effective bulk modulus $k$ defined by the following equation for both configurations:
\be
\varepsilon(Q_{\rm p}) = \frac{2\pi l^2}{v(Q_{\rm p})+2\pi l^2k}.
\label{eq:16}
\ee
This $k$ is plotted in Fig.2.
This plot clearly shows that even in the limit of $Q_{\rm p} \to 0$, $k$ for the parallel configuration is larger than that of the perpendicular configuration, and the energy gain is larger in the latter case.
\begin{figure}
\psfig{figure=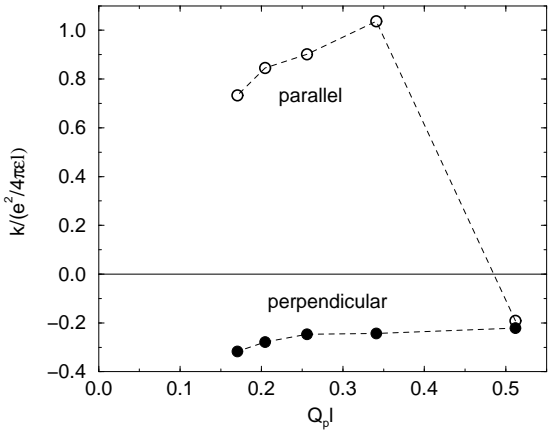,height=6.5cm}
\caption{The effective bulk modulus $k$ is plotted versus $Q_{\rm p}l$.
The dashed lines are guide to the eyes.
The open and closed circles are for the parallel and perpendicular configurations, respectively.}
\label{fig:2}
\end{figure}

Now we explain how the hydrodynamic theory gives the opposite result.
When the stripe runs parallel to the $x$-axis, or ${\vec Q}_{\rm s}$ is in the $y$-direction, the hydrodynamic theory describes the deformation of the stripe by the following Hamiltonian.~\cite{Fog,Mac,Lop}
\beq
H&=&\frac{1}{2L^2}\sum_{\vec q} \{(Yq_y^2+Kq_x^4)|u_{\vec q}|^2\cr
&+& [v({q})+\chi^{-1}]|\delta n_{\vec q}|^2 +2{\rm i}m_{\rm e}C q_y u_{\vec q}
\delta n_{-{\vec q}} \},
\eeq
where, $u_{\vec q}$ gives the displacement of the stripe, $\delta n_{\vec q}$ is the coarse-grained component of the electron density modulation due to the width change of the stripe, $\chi$ is the compressibility from which the effect of the Coulomb interaction is removed, $v(q)$ is the Coulomb potential, eq.(\ref{eq:2}), $m_{\rm e}$ is the electron mass, and $Y$, $K$, $C$ are parameters.
The compressibility is related to the ground state energy per electron, $E$, by the following equation:
\be
\chi^{-1} = 2\pi \l^2\frac{{\rm d}^2 (\nu E)}{{\rm d} \nu^2}.
\label{eq:13}
\ee
If $q$ is small we can equate $\delta n_{\vec q}$ and $\rho({\vec q})L_2(q^2l^2/2)$.
By adding eq.(\ref{eq:5}) to this Hamiltonian, we obtain the energy gain in the presence of the external potential as follows:
When the wave vector of the potential is ${\vec Q}_{\rm p}=(Q_{\rm p},0) \perp {\vec Q}_{\rm s}$,
\be
\delta E= -\frac{1}{2}\frac{2\pi l^2}{v(Q_{\rm p})+\chi^{-1}} V_{\rm ext}^2,
\label{eq:perp}
\ee
and if ${\vec Q}_{\rm p}=(0,Q_{\rm p})\parallel {\vec Q}_{\rm s}$,
\be
\delta E= -\frac{1}{2}\frac{2\pi l^2}{v(Q_{\rm p})+\chi^{-1}-(m^2C^2/Y)} V_{\rm ext}^2.
\ee
Because of the coupling between the displacement and width modulation of the stripes, the latter has lower energy.~\cite{note1}
This is opposite to our result.
The compressibility is estimated from the ground state energy to be $\chi^{-1}/2\pi l^2 \simeq -0.32e^2/4\pi\epsilon l$.
This value is quite close to the effective bulk modulus $k$ for the perpendicular configuration shown in Fig.2.
Thus our result agrees with that of the hydrodynamic theory only in the perpendicular configuration.

The origin of the present behavior is the charge modulation along the stripes, which is correlated between stripes.
To understand it, we first consider what happens when the filling factor is changed.
There are two orthogonal ways to change the filling factor.
In one way the width of the stripe is changed at fixed periodicity and in the other the periodicity is changed at fixed width of the stripes.~\cite{note3}
If the filling factor is changed uniformly, width change at almost fixed periodicity occurs.
This is because the stripe periodicity is determined by the minimum of the Hartree-Fock potential.~\cite{Kou1}
Next we consider unidirectional compression of the stripes.
If there is no charge transfer between the stripes, the compressibility will depend on the direction.
If they are compressed along the stripe direction, the periodicity is kept constant, and the width of each stripe changes.
This is easy direction, and the compressibility is high.
On the other hand, in the perpendicular direction the period changes and the width is kept constant.
This is hard direction, and compressibility is low.
In the presence of charge transfer between stripes, this kind of compression will never be realized.

However, the situation is different if periodic potential modulation of finite wavelength is applied to the present stripe phase, which is actually a rectangular CDW state with charge modulation along the stripes.
The charge modulation shows the tendency toward the Wigner crystallization, and the period is inversely proportional to the stripe width.
Inter-stripe correlation favors the stripes to have the same width in the direction perpendicular to the stripe.
In the perpendicular configuration, ${\vec Q_{\rm p}} \perp {\vec Q_{\rm s}}$, this correlation does not interfere the stripe to respond to $V_{\rm ext}$: the stripe changes the width as shown in Fig.1(b).
Thus the bulk modulus in eq.(\ref{eq:16}) is approximately given by $\chi^{-1}/2\pi l^2$.
On the other hand, in the parallel configuration, ${\vec Q_{\rm p}} \parallel {\vec Q_{\rm s}}$, stripes try to respond to $V_{\rm ext}$ keeping the width the same as neighbors.
So the main response is the displacement of their positions, and the periodicity is modulated.
The energy gain is smaller in this situation.
Actually, the bulk modulus of the period change is estimated to be $k \simeq 1e^2/4\pi\epsilon l$, which is consistent with Fig.2.
The reason that $k$ decreases as $Q_{\rm p}$ decreases is that slight width modulation is allowed at longer wavelength.

The reason why $m=2$ case has low $k$ even for the parallel configuration is explained by this scenario.
In this case the displacement of the electron-stripe means the width change of the hole-stripe at fixed periodicity.
This is unique to $m=2$, since for $m\ge 3$ the displacement of the electron-stripes causes both the displacement and width change of the hole-stripes.
Therefore, only $m=2$ has low bulk modulus in the parallel configuration.

From the argument above we propose to modify the hydrodynamic theory such that the compressibility $\chi$ has ${\vec q}$ dependence.
Another fact ignored in the hydrodynamic theory is that the displacement of the stripe induces local charge modulation as pointed out above.
Namely, density modulation $\delta n_{\vec q}={\rm i}q_yn_0u_{\vec q}$ is induced by $u_{\vec q}$, where $n_0$ is mean electron density.
Hydrodynamic theory including these effects should be constructed.

Ishikawa and Maeda also calculated the effect of the external potential.~\cite{Ishi}
They applied the second order perturbation theory to their Hartree-Fock ground state.
Their conclusion is that there is no energy gain in the parallel configuration, and the perpendicular configuration has lower energy.
Although the qualitative conclusion is the same as ours, we doubt validity of their calculation.
As we have shown here the coupling to the external potential always deforms the stripe state, and lowers the energy of the system.
The lack of the energy gain in their parallel case indicates that the intermediate states are not correctly taken into account.
Thus their conclusion that the perpendicular configuration is favored has no foundation.

In conclusion we have shown that the stripes orient themselves perpendicular to the long wavelength external modulation to gain energy without loosing correlation intrinsic to the stripe phase.

Part of the numerical calculation is performed at the ISSP, the Univ. of Tokyo.
This work is supported by Grant-in-Aid No.~12640308 from MEXT, Japan.

\vfill\eject
\begin{table}
\caption{The value of $\varepsilon e^2/4\pi \epsilon l$, which is the coefficient of the energy gain per electron in the external potential. The size of the wave vector of the external potential is $Q_{\rm p}=Q_{\rm s}/m$, the direction of which is either parallel or perpendicular to the stripe's wave vector.}
\label{table:1}
\begin{tabular}{cccc}
m & $Q_{\rm p}l$ & parallel  & perpendicular \\
\tableline
2 & 0.5117     & 0.568  & 0.577  \\
3 & 0.3411     & 0.252  & 0.372  \\
4 & 0.2558     & 0.208  & 0.273  \\
5 & 0.2047     & 0.175  & 0.217  \\
6 & 0.1706     & 0.152  & 0.180  \\  \hline
\end{tabular}
\end{table}


\begin{references}
\bibitem{note2} When we increase the strength of the external potential further in the perpendicular configuration, the modulation in the width increases further, and finally the stripes are connected at the bulged regions.
When $V_{\rm ext}$ becomes larger than 0.3$e^2/4\pi\epsilon l$, the pattern no longer looks like stripe, but looks more like square CDW state.
\bibitem{note1} The parameter $Y$ must be positive for the system to be stable.
\bibitem{note3} It should be noticed that in the latter case both the width and the periodicity of the hole-stripes change.
\end{references}
\end{document}